# Beyond second-long trajectory of the Trp-cage peptide generated using a Kinetic Monte Carlo model derived from molecular dynamics

Abhijit Chatterjee, Rishav Deb, Gauri Thapa and Swati Bhattacharya*

Department of Chemical Engineering, Indian Institute of Technology Bombay, Mumbai 400076, India

*Email: swaticb@che.iitb.ac.in

## Abstract

We present a kinetic Monte Carlo (KMC) modeling approach to describe the stochastic dynamics of a peptide molecule spanning nanosecond to second timescales. The dynamics of protein conformational changes is interpreted at a local level in terms of dihedral transitions. Taking Trp-cage miniprotein as an example, the KMC model "learns" about the transitions from multiple MD trajectories. Training is based on local divide-and-conquer strategy that identifies the discretized backbone dihedral states as building blocks for the conformational space, along with associated transition rates of dihedral flips to describe the conformational state-to-state dynamics. A key feature in our approach is the incorporation of backbone correlations, such that rates are conditioned on the local environment and steric coupling. We show that with the correlations built-in, the KMC model closely matches MD. Such an approach is shown to reach second timescales in a few CPU hours on a standard desktop computer, and can easily yield multiple stochastic realizations of the conformational dynamics. Our KMC model construction scheme should be generally applicable to a wide range of proteins, and can be used for bridging local flexibility to protein-wide dynamics.

Keywords: Kinetic Monte Carlo, Molecular dynamics, protein dynamics

## 1. Introduction

Protein function is related to both structure and dynamics. In many cases, dynamics is the real engine of the function[1,2]. Binding, catalysis, allostery, and signaling often depend on subtle rearrangements that unfold across multiple timescales and involve strongly coupled local and global motions. Despite the centrality of these processes, the full dynamical pathways by which proteins switch between functional states remain difficult to resolve both experimentally and computationally.

Molecular dynamics (MD) remains the gold-standard among computational approaches to study protein dynamics[3–6]. By following the local conformational rearrangements at atomistic scale resolution, one can better elucidate the large-scale dynamical changes in protein structure and function. However, MD simulations are computationally intensive especially when investigating long-time dynamics of a protein in the timescale of microseconds to milliseconds or beyond[7]. This is because atomistic MD employs a small timestep to resolve the fast vibrations of the protein atoms. Despite algorithmic[8–20], software[21–23] and hardware breakthroughs[24,25] as well as recent advances in artificial intelligence[26,27], accessing the long-time dynamics of proteins using MD still remains a grand challenge, especially for large proteins.

Atomistic kinetic Monte Carlo (KMC) approaches, which have been remarkably efficient in reaching long timescales in inorganic systems[28–35], remain relatively underdeveloped for protein systems due to the challenges of mapping continuous macromolecular configuration spaces onto discrete transition networks. While Monte Carlo (MC) methods have efficiently explored protein folding equilibria and landscapes since the 1990s—using force fields or Go-like potentials to study ensemble pathways (e.g., Shimada[36] et al., 2002; Heilmann et al.[37])— these approaches fundamentally lack true time-dependent kinetics. Peter et. al.[38–40] present a hybrid MD/KMC algorithm for protein folding and aggregation in explicit solvent, alternating short MD trajectories with discrete, timescale-bridging KMC moves. To accelerate rare events, their framework relies on isolated, hand-crafted move sets targeting specific secondary (e.g., dihedral rotations) or tertiary (e.g., translational displacements) structural transitions. Consequently, the method is not easily transferable to complex or novel biomolecular systems where these state-to-state pathways are unknown *a priori*.

Against this background, we present a dihedral-resolved KMC framework that “learns” the slow local torsional state transitions from MD trajectories, as well as the backbone residue correlations, and the resulting KMC rate catalog is used to propagate the dynamics. We model the protein conformational dynamics using dihedral flips. MD simulations provide time series data for building the KMC model. Once the KMC model is constructed, it can be used to generate new synthetic trajectories in terms of discrete dihedral states, from which one can reconstruct the full molecular structure/dynamics.

Key features of our KMC model is the incorporation of backbone correlations, environment-conditioned transitions where single-dihedral jump rates depend on neighboring configurations, and concerted transitions where adjacent dihedrals undergo synchronous moves. We show that a KMC model lacking these features would provide a dynamical behavior significantly different from MD. The proposed method provides protein structure dynamics while being more than a million times faster than atomistic MD.

The development of atomistic dihedral-level KMC model also fills an important gap between MD and commonly-used low-dimensional or coarse-grained kinetic models that employ a limited number of collective variables, order parameters, or low-dimensional reaction coordinates[41–45] to focus on the essential dynamics during protein structural rearrangements.

## 2. Methods and Materials

We first describe the MD protocols before delineating the KMC model.

### 2.1 System Preparation for MD simulations

The Trp-cage miniprotein is selected as the example for this study. Trp-cage has been widely studied both experimentally as well as for testing enhanced sampling methods[46–49]. The initial structure of the Trp-cage peptide (sequence DAYAQWLKDGGPSSGRPPPS) was obtained from the NMR structure (PDB ID 2JOF[50]), as illustrated in Figure 1(a). To solvate the system, the peptide was placed in a cubic water box of dimension $6 \times 6 \times 6\ nm^3$ consisting of $\sim 7050$ TIP3P[51] water molecules. The final system was charge-neutral, comprising roughly of 21,500 atoms in total.

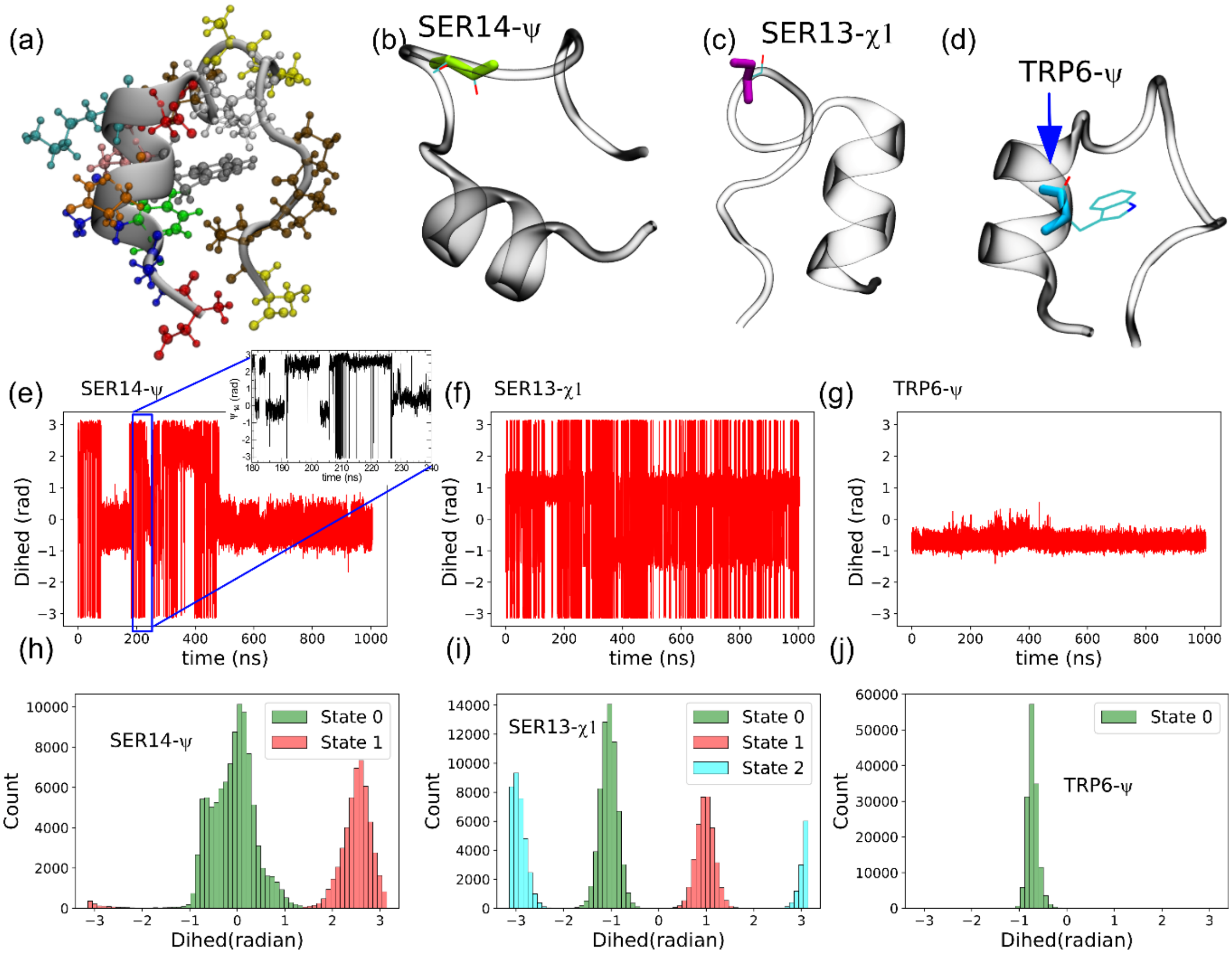

Figure 1. (a) Snapshot of the Trp-cage protein in ribbon representation (backbone), with side-chain atoms depicted as ball-and-stick models colored by residue type. (b), (c) and (d) represent the $\psi$ dihedral of Ser14, the $\chi 1$ dihedral of Ser13 and the $\psi$ dihedral of Trp 6 in stick representation. The side-chain of the respective residue is also indicated by lines. (e), (f) and (g) The time-series of the three dihedrals obtained from a 1 $\mu s$ MD trajectory. Inset in (e) shows a magnified view of the fluctuations in $\psi_{14}$ between 180 and 240 ns. (h), (i) and (j) Histograms of the dihedrals $\psi_{14}$, $\chi 1_{13}$ and $\psi_6$ (the subscripts representing the residue index of the two middle atoms of the dihedral) constructed from the MD data. The peaks, identified as the discrete states are color coded.

## 2.2 MD Simulation Protocols

All MD simulations were carried out using the NAMD 2.14 package[21,52] with CHARMM36 force fields[53,54]. Periodic boundary conditions were applied and long-ranged electrostatic interactions were computed using the Particle-mesh Ewald method[55]. Covalent bonds containing hydrogen atoms were constrained using the SETTLE[56] and RATTLE[57] algorithms, allowing an integration time step of 2 fs. Non-bonded interactions were calculated using a cutoff distance of 12 Å with a switching distance of 10 Å. Energy minimization was initially performed for 3,000 steps of conjugate gradient. The system was subsequently equilibrated in the NPT ensemble at 310 K for a minimum of 5 $ns$ using a Nosé–Hoover Langevin piston[58] for pressure regulation. Final MD production runs were conducted in the NVT ensemble at 310 K, controlled by a Langevin thermostat. Ten independent trajectories of varying lengths, with diverse initial conformations given in the Supporting Information Table S1, were obtained with frames recorded every 10 $ps$. The cumulative trajectory length was 7.3 $\mu s$. The protocols used for generating the initial conformations is given in the Supporting Material Section S1.

### 2.3 Reconstruction of peptide conformations

Peptide structures predicted by KMC methods are reconstructed using PeptideBuilder[59] based on the backbone dihedral values.

## 3. Results and Discussion

We first present important dynamical insights gained from MD relevant to the construction of the KMC model.

### 3.1 From MD trajectories to discrete dihedral state timeseries

It is possible to reconstruct the protein structure atom-by-atom from the given bond lengths, angles and dihedrals. Since the bonds and angles usually exhibit narrow, unimodal distributions, the conformational diversity of a protein arises from the dihedral variations. Thus, protein dynamics can be followed in terms of the dihedral angle dynamics. We begin with an analysis of typical dihedral distributions obtained from MD calculations.

By focusing on individual dihedrals, one can perform modular piece-wise analysis of peptide. Three different dihedrals are presented in Figure 1 (b), (c) and (d) as examples, viz. $\psi$ dihedral of Ser14, $\chi 1$ dihedral of Ser13 and $\psi$ dihedral of Trp6.

Figure 1(e), (f) and (g) shows the variation of the three dihedrals computed from a $1\mu s$ MD trajectory. The histograms for the corresponding dihedrals are shown in Figure 1(h),(i) and (j). Broadly, each dihedral shows a small number of distinct states identified by the color-coded peaks in the histograms (Figure 1h,I,j). The $\psi_6$ (the subscript refers to the residue index of the two middle atoms of the dihedral) dihedral has only one state indicated by the unimodal distribution (Figure 1j). The $\chi 1_{13}$ dihedral (a side-chain rotamer of Ser13) shows three states, with one of the states wrapped around the periodic boundary (Figure 1i). $\psi_{13}$ has at least two peaks as indicated in the histogram in Figure 1(h). There may be multiple overlapping states in case of $\psi_{13}$, but these are not readily resolved.

Figure 1 (e)-(g) indicate that a dihedral involves of two types of motion. The first is the fast vibration within a state. The second is the excursion from one state to another, which occurs quite infrequently relative to the vibrations. Inter-state transitions happen instantly, as an abrupt jump and not as a slow, gradual drift. This is indicated by the inset in Figure 1(e) which shows a magnified image of a smaller segment of the trajectory between 180 and 240 ns. In this example, the dihedral $\psi_{14}$ intermittently visits two separate regions that may be identified as the local states while also showing vibrations within each state. In case of $\psi_{14}$, a large number of such inter-state transitions happen at the microsecond timescale.

These observations, which generalize across all backbone dihedrals in the Trp-cage protein, highlight several key characteristics of the system's dynamics. Backbone motion can be discretized into distinct conformational states defined by individual dihedral angles $(\phi, \psi)$, with transition rates extracted directly from MD simulations. We assume that dihedral dynamics entails well-defined state-to-state transitions of Markovian nature. Furthermore, we assume rapid intra-state relaxation, neglecting vibrational fluctuations within energy wells. This fundamental separation of timescales between fast vibrational motion and state-to-state transitions is illustrated in Figure 1(e) and its inset.

### 3.2 Correlations between dihedrals

Another aspect revealed by MD is that the backbone dihedrals can being dynamically correlated, i.e., the joint distribution of dihedral states cannot be obtained in terms of the single dihedral distributions, and the transition rates are conditional on the neighbor dihedral state(s). To determine the spatial extent of local correlations affecting dihedral states and transition rates, we investigate both the joint probability distributions and the collective dynamics of adjacent backbone dihedrals.

The Trp-cage miniprotein consists of an N-terminal $\alpha$-helix, a central loop region, and a C-terminal polyproline stretch. We focus on a specific dihedral, $\phi_{11}$, located within the central loop of the Trp-cage miniprotein—a particularly labile region of the peptide. First consider $\phi_{11}$ along with its right neighbor, $\psi_{11}$. $\phi_{11}$ possesses a bimodal distribution whereas $\psi_{11}$ has a trimodal distribution (see the dihedral histograms in Figure 2(a) and (b)). The MD-derived joint probability density map for GLY11 (Figure 2(c)) reveals discrete, high-density basins rather than a continuous distribution. This energy landscape supports modelling the dihedral dynamics as discrete state-to-state transitions (hopping between basins).

To assess the dihedral coupling, a synthetic joint map is constructed from the product of the individual marginal histograms under the assumption of independence (Figure 2(d)), serving as an uncorrelated baseline, i.e., $P(\phi_{11}, \psi_{11}) = P(\phi_{11})P(\psi_{11})$. The fact that the two dihedrals are correlated is clearly demonstrated by the departure of the MD-derived map from this baseline—specifically through the shifting, fading and intensification of energy basins in Figure 2(c) compared to panel (d).

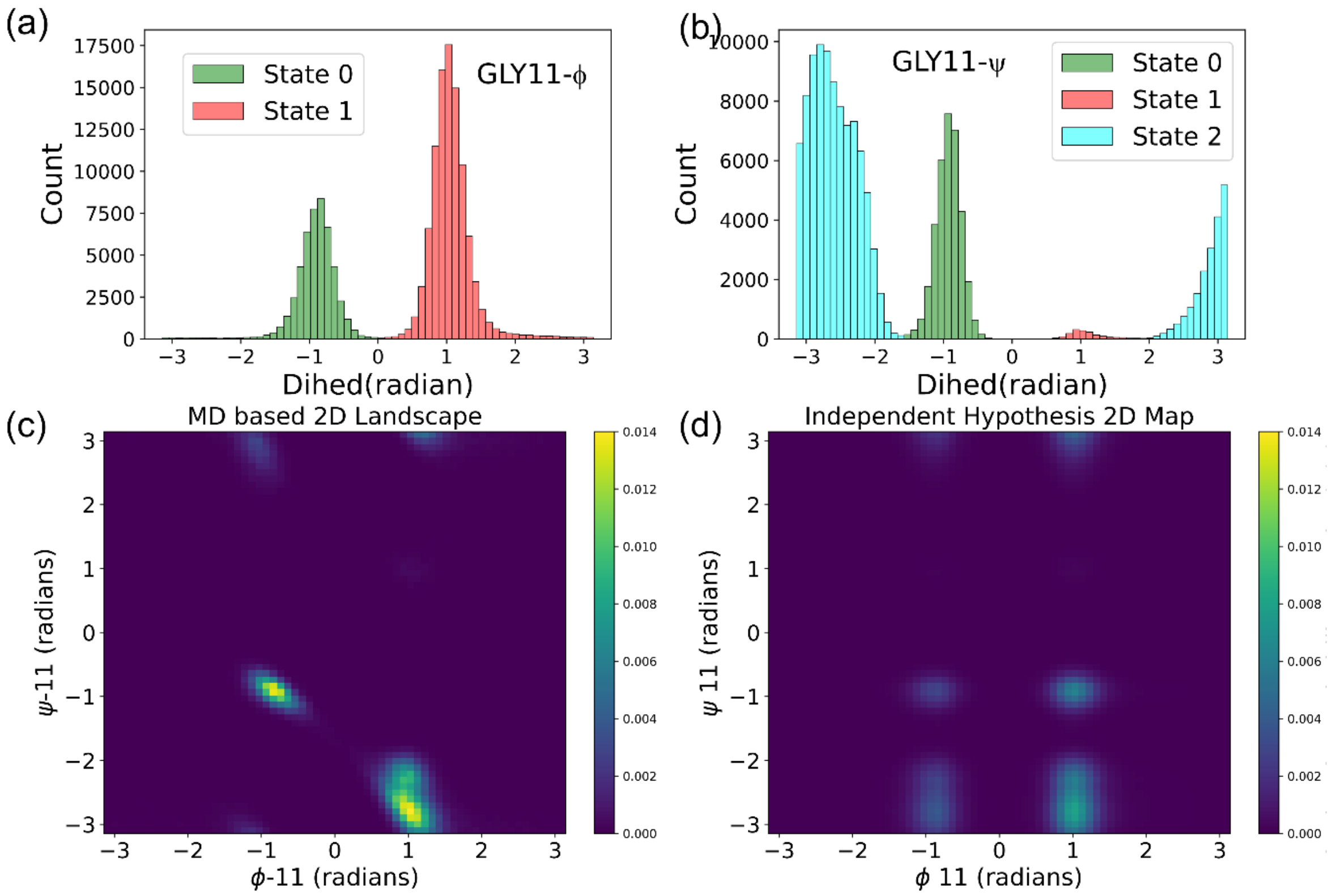

Figure 2 (a)-(b) Histograms for the dihedrals $\phi_{11}$ and $\psi_{11}$ respectively, obtained from the MD data. (c) 2D map showing the joint probability density of $\phi_{11} - \psi_{11}$ from MD. (d) Synthetic map generated for the joint probability density of $\phi_{11} - \psi_{11}$ from the single dihedral distributions assuming that the dihedrals are mutually independent ($P(\phi_{11}, \psi_{11}) = P(\phi_{11}) * P(\psi_11)$).

Expanding on this analysis, Figure 3 illustrates how these dependencies change with increasing sequence distance. Figure 3(a) shows clear coupling between the central dihedral, $\phi_{11}$ and its immediate left neighbor, $\psi_{10}$, highlighted by significant structural differences from the synthetic map (Figure 3(b)). However, as shown for second and third neighbors in Figure 3(c–f) as well as the Supporting Material Figure S1, the MD-derived distributions progressively resemble their synthetic baselines, indicating that correlations rapidly diminish with inter-dihedral distance.

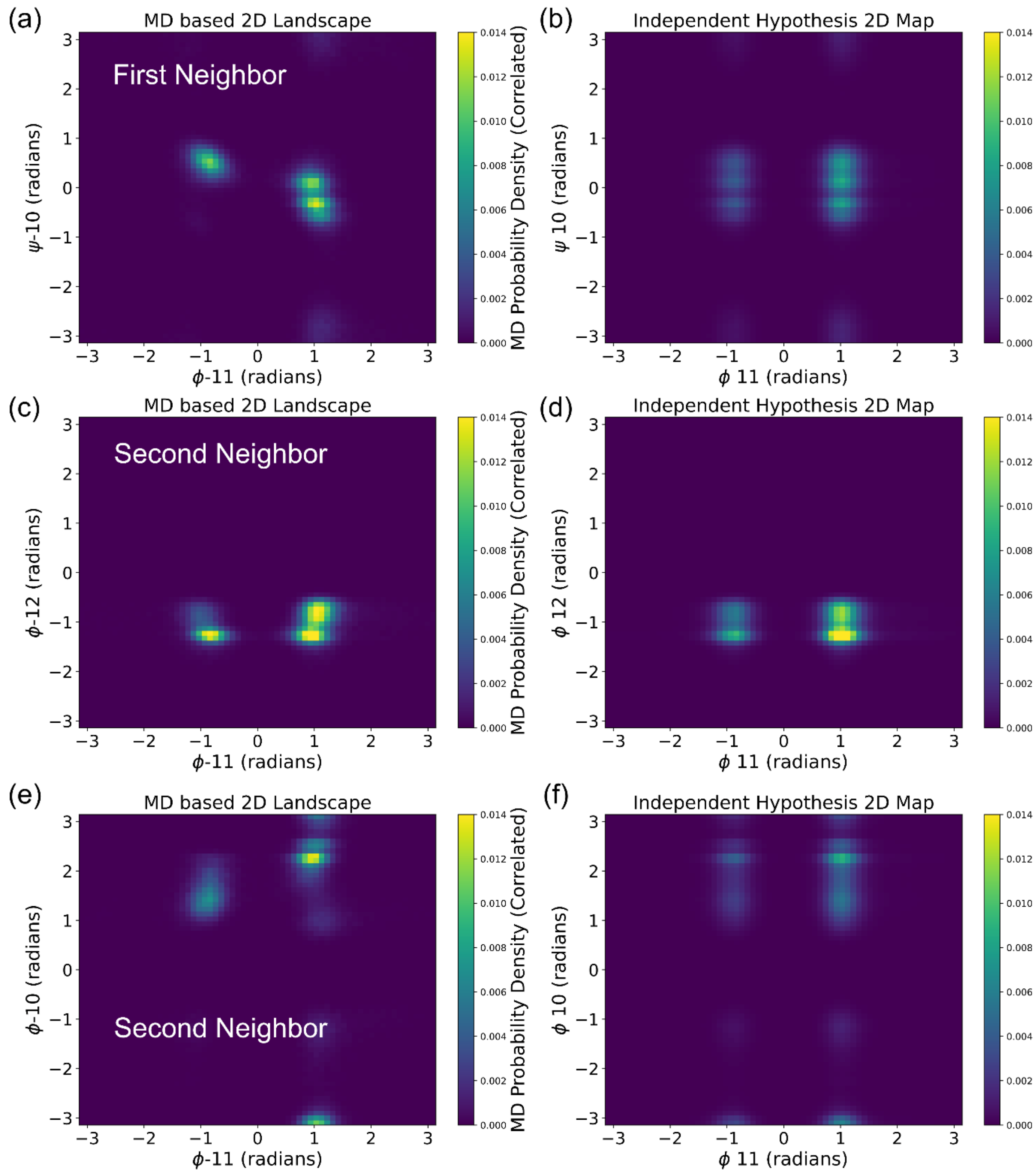

Figure 3. Comparison of MD-derived (left panels) and synthetic (right panels) joint probability maps for neighboring dihedrals. Panels on the left show MD derived joint probability maps of $\phi_{11}$ with neighboring dihedrals. Panels on the right show the corresponding maps generated synthetically from the respective marginal distributions assuming that the dihedrals are independent. (a), (b) MD derived and synthetic joint probability maps of $\phi_{11} - \psi_{10}$. (c),(d) MD derived and synthetic maps of $\phi_{11} - \phi_{12}$, which are second neighbors. (e),(f) MD derived and synthetic maps of $\phi_{11} - \phi_{10}$ .

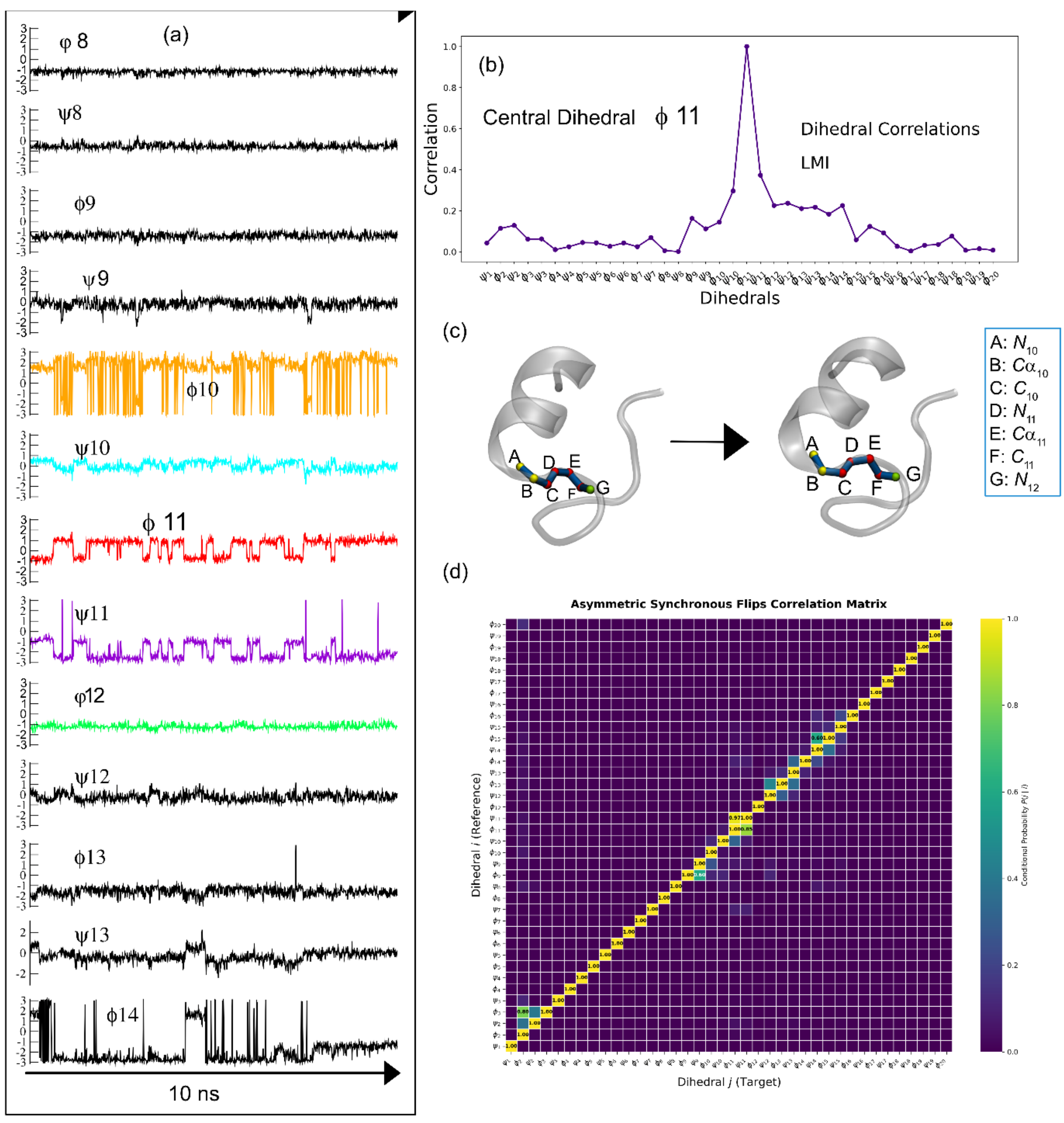

Figure 4. (a) MD derived time-series profiles for the dihedrals $\phi_8$ to $\phi_{14}$ over a 10 ns fragment of a trajectory. The time-series for the dihedrals are stacked according to their sequence in the peptide backbone. (b) Correlation profile of $\phi_{11}$ with all other backbone dihedrals,

calculated using Linear Mutual Information (LMI). (c) Conformational snapshots capturing a transition within consecutive backbone dihedrals ($\psi_{10}, \phi_{11}$ and $\psi_{11}$). The three successive dihedrals are represented as sticks, with their constituent atoms depicted as labeled spheres ($A$ through $G$). Specifically, $\phi_{11}$ is defined by atoms $C\ D\ E\ F$ (with atom names indicated adjacent to the spheres), $\psi_{10}$ by atoms A B C D, and $\psi_{11}$ by atoms $D\ E\ F\ G$. The snapshots represent two consecutive frames in a trajectory at the point of the transition. (d) Map showing fraction of synchronous moves by pairs of backbone dihedrals.

Figure 4(a) illustrates a 10 ns trajectory segment tracking the sequential dihedrals ($\phi_8$ to $\phi_{14}$). The purpose of the stacking of the time-series is to visually identify co-moving dihedrals, if any. The first four dihedrals ($\phi_8 - \psi_9$) from the top, which correspond to the $\alpha - helical$ region of the peptide, show limited vibrational motion without distinct flips. The dihedrals $\phi_{10}$ to $\psi_{11}$ show frequent flips that may be identified as inter-state transitions. The next three dihedrals, $\phi_{12}$ to $\phi_{13}$, exhibit only vibrational motion within the time interval. The last two dihedrals (especially $\phi_{14}$) show a number of jumps in the 10 ns. Thus, zooming in on the central dihedral $\phi_{11}$, demonstrate simultaneous coordinated jumps with its right neighbor ($\psi_{11}$); the strong kinetic coupling between $\phi_{11}$ and $\psi_{11}$ is physically grounded by their three shared backbone atoms. Structural snapshots capturing this concerted transition at 6.4 ns from the start of the trajectory segment, are shown in Figure 4(c). However, the left neighbor ($\psi_{10}$) exhibits weaker coupling; its time series reveals poorly resolved states with overlapping distributions, complicating discrete assignment.

To quantify this spatial decay in correlations, we calculated the Linear Mutual Information (LMI) between $\phi_{11}$ and all remaining backbone dihedrals (Figure 4(b)). The resulting correlation profile confirms that significant correlations are restricted to immediate spatial neighbors and decay precipitously with sequence distance. Notably, the mutual information between $\phi_{11}$ and its second-nearest neighbors ($\phi_{12}$ and $\phi_{10}$) drops to baseline noise levels characteristic of distant residues. These data justify implementing a strict nearest-neighbor cutoff when defining the localized environmental context for each dihedral variable.

Next, we examine the underlying nature of these dihedral correlations. Inter-dihedral coupling can manifest through distinct mechanisms: the conformational state of one dihedral may alter the transition dynamics of another, or transitions across neighboring dihedrals may occur synchronously. The time-series trajectories in Figure 4(a) indicate that adjacent dihedrals occasionally undergo concerted transitions, a phenomenon most pronounced between $\phi_{11}$ and $\psi_{11}$. To quantify this effect, we evaluate the frequency of synchronous transitions across all dihedral pairs, including non-adjacent pairs, and represented the results as a pairwise synchronicity matrix (Figure 4(d)). For each pair of dihedral $(i, j)$ in a given order, the synchronicity matrix gives the fraction of moves of dihedral $i$ that are synchronous with moves of $j$. Two dihedrals $j$ and $j$ with synchronized moves, can also independently participate in other moves. For example, consider two dihedrals $A$ and $B$ that exhibit concerted moves. Now, it is possible that for $A$, 90% of its moves coincide with moves of $B$. And thus, only 10% of the moves of $A$ can be deemed as solitary (or strictly single dihedral moves). On the other hand, $B$ may have a much smaller fraction of moves (say, 70%) that are synchronous with $A$, i.e. the two dihedrals can have different numbers of solitary moves. As a result, the synchronicity matrix is, in general, asymmetric. While high synchronicity is rare across the system, a small subset of dihedral pairs exhibits concerted transitions that must be explicitly accounted for in the KMC model.

### 3.3 State Space Discretization

A key goal is to identify define states of a dihedral from its distribution. Here, we apply the cluster-by-valley method, as described below. Figure S2 in the Supporting Material shows selected MD-derived dihedral distributions with the colored peaks indicating the algorithm-identified states. In cases where peaks are well separated, it is straightforward to associate one peak with a state. Overlapping peaks in $\psi_{10}$ and $\phi_{10}$ distributions in Figure S2 can become challenging to resolve. Such broad peaks, which indicate considerable conformational flexibility, serve an important function. While sterically coupled adjacent dihedrals sharing multiple atoms are expected to undergo concerted flips, this localized flexibility allows neighboring dihedrals to accommodate a structural transition through subtle vibrational adjustments rather than full, synchronized reorientations.

We computed the time series for all 38 backbone ($\phi, \psi$) dihedrals from MD trajectories and discretized each using an unsupervised cluster-by-valley method[60]. For each dihedral, we generated a density from one-fifth of the cumulative trajectory data via kernel density estimation (Gaussian kernel with 0.3 rad bandwidth) over the periodic domain [-π, π). We identified multi-modal peaks/valleys, defining basin boundaries at valleys below a valley-ratio threshold (0.7) relative to adjacent peaks.

The circular boundary condition requires additional care. This is evident in the histograms shown for example dihedrals in Figure 1(h), Figure 2 (b) and Figure S2 in the Supporting Material. In case of $\psi_{11}$ (i.e., $\psi$ dihedral of GLY11), the two peaks at the boundaries correspond to a single state. Both are shown in cyan color in the histogram in Figure 2b. We merge periodic peaks separated by shallow valleys (and within a small circular distance) and filter clusters by minimum occupancy to remove transients. The resulting states (typically between 1-4 per dihedral, and labeled as dihedral state 0, 1, 2, ...) are characterized by boundary angles, circular mean centers, and von Mises concentration (yielding angular σ).

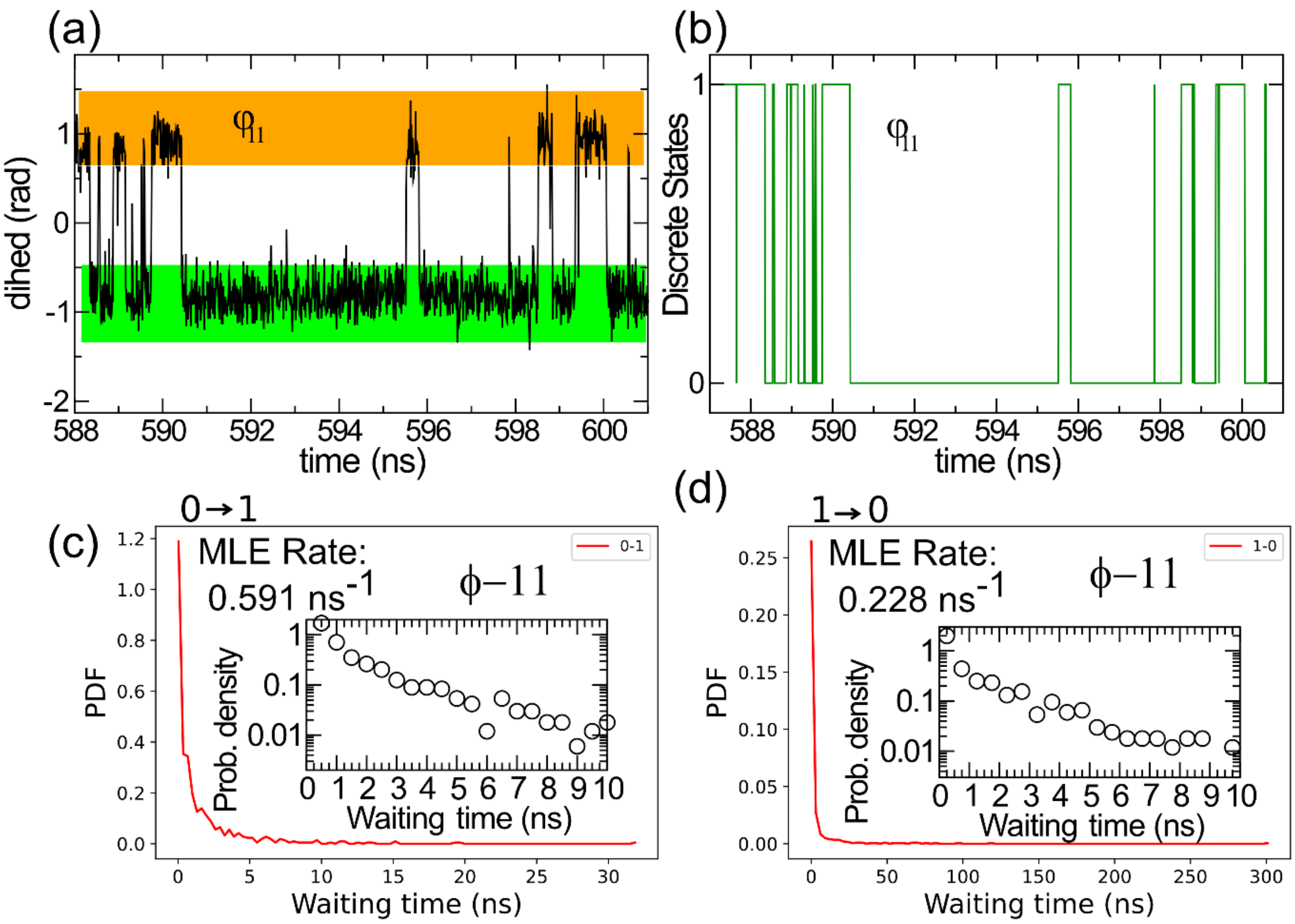

Figure 5. (a) Variations in the dihedral $\phi_{11}$ over a 14 ns segment of an MD trajectory. The two colored bands indicate the two discrete states identified by the clustering algorithm. (b) Discrete state time series for $\phi_{11}$ for the same portion of the trajectory shown in (a). The two states are labelled 0 and 1. (c) and (d) Waiting time probability density for transitions of $\phi_{11}$ from state 0 to 1 and from 1 to 0 respectively. The insets show the waiting time probability density in semi-log scales till 10 ns.

Once this procedure identifies dihedral states, we transform the continuous dihedral angle-time series data into a discrete dihedral state one. Figure 5(a) and (b) illustrate such a transformation for $\phi_{11}$. The snapshots in Figure 4(c) are representative images of two different states of the dihedral $\phi_{11}$ discovered through the clustering algorithm. The dihedral time series for $\phi_{11}$ from a segment of an MD trajectories is shown in Figure 5(a) with the two states indicated by colored bands. Figure 5(b) shows the corresponding discrete state time series for the same dihedral, $\phi_{11}$. The dihedral flips between the two states, labelled 0 and 1.

This discretization procedure was applied independently to each dihedral angle. Notably, $Gly10 - \phi$ and $Gly11 - \phi$ possess distinct states defined by their own individual time series, even though both belong to the same amino acid residue type.

### 3.4 Estimating transition rates for KMC

The discrete state time-series in Figure 5(b) shows the dihedral flip between states 0 and 1 several times. The waiting time probability densities for the same dihedral in panels (e), and (f) reveals exponential distribution, as confirmed by the semi-log insets displaying the same data, corresponding to a first-order process. We use the data from multiple MD trajectories to calculate the transition rates for the moves using the maximum likelihood estimate (MLE).

The rate constant, $k_{i\to j}$, for the transition from state $i$ to $j$ is estimated as $k_{i\to j} = N_{i\to j}/t_i$, where $N_{i\to j}$ is the total number of transitions from $i$ to $j$, and $t_i$ is the total waiting time in state $i$, which includes the waiting times for transitions from state $i$ to other states as well. Here, $N_{i\to j}$ and $t_i$ are computed using the accrued data from all trajectories. In order to avoid recrossing events, only transitions where the dihedral remained in the final state for at least 40 $ps$ are considered. Thus, rapid transitions or flickering were filtered out prior to the calculation of transition rates.

To incorporate dihedral correlation discussed earlier, this estimator is extended across three distinct move types:

1. Context-Conditioned Moves**:** For an individual dihedral $A$ undergoing a transition from state $i$ to $j$, whose rate is conditioned on the current state of its neighbors, we use the term context $c$, which is defined as the joint state of its two immediate flanking neighbors, $c = (s_{A-1}, s_{A+1})$. The rate is estimated as $k_{i\to j|c} = \frac{N_{i\to j|c}}{t_{i|c}}$, where $N_{i|c}$ is the number of $i$ to $i$ flips observed in context $c$, and $t_{i|c}$ is the total residence time in state $i$ under the same context.

2. Synchronized Moves: For a pair of dihedrals $(A, B)$ undergoing simultaneous transition $(i_A, i_B)$ to $(j_A, j_B)$ conditioned on their combined context $C$, the joint rate is estimated as $k_{(i_A,i_B)\to(j_A,j_B)|C} = \frac{N_{(i_A,i_B)\to(j_A,j_B)|C}}{t_{(i_A,i_B)|C}}$, where $N_{(i_A,i_B)\to(j_A,j_B)|C}$ denotes the number of synchronous flips observed, and $t_{(i_A,i_B)|C}$ is the dwell time in state $(i_A, i_B)$ under context $C$.
3. Solitary Moves: For dihedrals capable of synchronous motion, solitary transitions (eg. Dihedral $A$ flipping from $i_A$ to $j_A$) while $B$ remains unchanged are parameterized as $k_{i_A\to j_A|solitary} = \frac{N^{solitary}_{i_A\to j_A|c}}{t_i|c}$where $N^{solitary}_{i_A\to j_A|c}$ strictly counts flips verified to occur independently in dihedral $A$ without a concurrent flip in dihedral $B$.

Note that some KMC models described next may not include move types 2 and 3.

## 3.5 Hierarchy of KMC models

A hierarchy of KMC models is developed in this study. The role of KMC is to generate discrete timeseries data for the entire Trp-cage backbone analogous to Figure 5(b). For each model, one hundred independent KMC runs were spawned starting from different initial conformations (sets of dihedral states) selected randomly from the MD trajectories. Each trajectory was terminated when the KMC clock exceeded 1 $\mu s$.

### 3.5.1 Independent KMC model- Model A

In the simplest model, dihedrals evolve independently via the Gillespie algorithm[61,62], and contexts and synchronized moves are ignored. Only the initial and final state of a dihedral, and the associated rate is considered. At each step, a specific transition channel $\mu$ (a particular move $i$ to $j$ of a given dihedral) is selected from all available moves with a probability proportional to its individual rate $k_\mu$ relative to the total escape rate $k_{tot} = \sum_m k_m$ , such that $\sum_{m=1}^{\mu-1} k_m < u_1 k_{tot} \leq \sum_{m=1}^{\mu} k_m$ using a uniform random number $u_1 \in (0,1]$. The simulation clock then advances by $\Delta t = -\frac{-\ln(u_2)}{k_{tot}}$ using a second independent random number $u_2 \in (0,1]$, the state of the selected dihedral is updated, and the rate catalog is re-evaluated for the next step.

### 3.5.2 Context-conditioned KMC model- Model B

In the correlated KMC model, one incorporates context-dependent transition rates. The simulation begins with the initialization of all dihedrals using a randomly selected MD conformation. At each Gillespie step, the algorithm evaluates the current local state of each dihedral and constructs its real-time context $c = (s_{A-1}, s_{A+1})$ based on its immediate flanking neighbors. All permitted context-conditioned transitions across all dihedrals form the active system-wide rate list with a total escape rate $k_{tot} = \sum_m k_{m|c}$. A specific transition channel $\mu$ is selected with probability proportional to its rate $\frac{k_{\mu|c}}{k_{tot}}$ using a uniform random number $u_1 \in (0,1]$. The KMC clock then advances by $\Delta t = -\frac{-\ln(u_2)}{k_{tot}}$ using a second independent random number $u_2$ and the state of the target dihedral is updated.

### 3.5.3 Context-conditioned KMC model with synchronized moves- Model C

In this model, the rate catalogue is expanded to explicitly account for synchronous dihedral transitions identified from the underlying MD trajectories. Two dihedrals are defined as synchronous if they undergo joint transitions within a 20 ps interval (equivalent to 2 frame steps). A dihedral pair is added to the synchronous catalogue if joint transitions constitute more than 50% of the total moves for at least one of the participating dihedrals, with a minimum threshold of 30 observed synchronized flips. Although non-adjacent synchronous events were occasionally detected in the MD trajectories, only immediate neighboring pairs satisfied these strict criteria. For all non-synchronized dihedrals, standard context-conditioned transitions are maintained. In the Trp-cage model, only two synchronous pairs of dihedrals were detected, namely, $(\phi_{11}, \psi_{11})$ and $(\psi_{14}, \phi_{15})$.

The simulation initializes all dihedrals using a randomly selected MD conformation. At each Gillespie step, the algorithm re-evaluates the local state of all dihedrals to construct their real-time contexts. The overall system rate list incorporates both the standard single-dihedral transitions and the specialized transition channels for each synchronous pair. These pair-specific channels comprise both joint moves conditioned on $C_{AB}$ and solitary moves as discussed earlier. The total escape rate is evaluated at each time step based on all allowed transitions. A specific transition channel $\mu$ is selected with probability proportional to its rate $\frac{k_{\mu|C}}{k_{tot}}$ using a uniform random number $u_1 \in (0,1]$. The KMC clock time then advances by $\Delta t = -\frac{-\ln(u_2)}{R_{tot}}$ using a second independent random number $u_2$, the state of the target dihedral is updated, and all affected neighbor contexts are re-evaluated for the subsequent step.

### 3.6 Comparing probability distributions from MD and KMC models

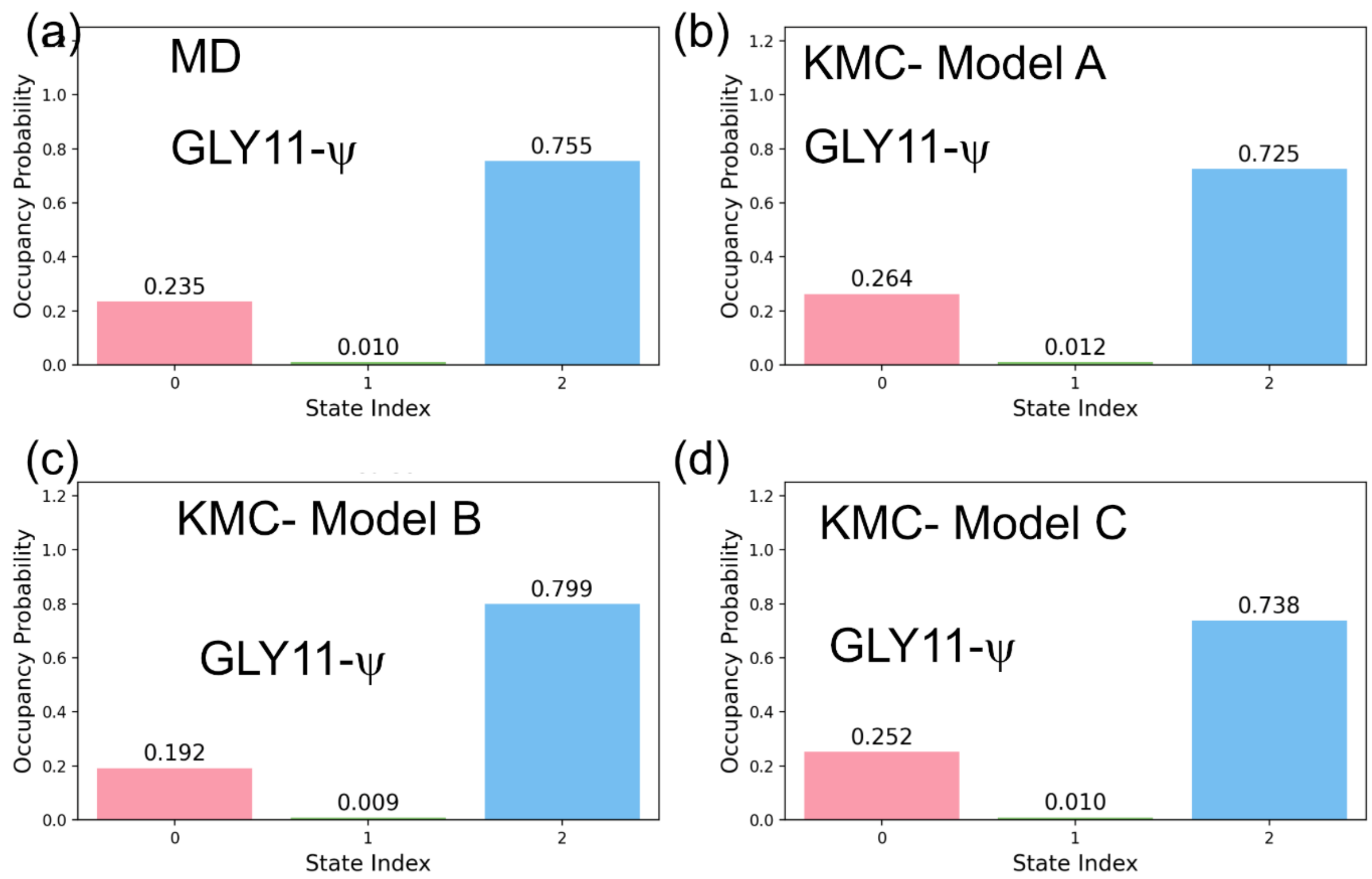

Figure 6. State occupancy probability of a selected dihedral, $\psi_{11}$ computed from the MD, Independent (Model A), Context-conditioned (Model B) and CC-Synced (Model C) KMC methods.

To evaluate model performance at the individual dihedral level, we compare the equilibrium state occupancy probabilities obtained from the original MD trajectories against those generated by the three KMC models of Section 3.5. All models reproduce reasonably well the MD (discretized) dihedral distributions across all dihedrals. Figure 6 displays the state probability distribution for a representative dihedral, $\psi_{11}$. Corresponding occupancy profiles for additional sample dihedrals ($\phi_{13}$ and $\phi_{11}$) are provided in Figures S3 and S4 of the Supporting Material.

In order to compare the state populations across all dihedrals using a single metric, we quantified differences using the mean KL (Kullback–Leibler) divergence, $\overline{\mathrm{KL}} =$ $\frac{1}{N}\sum_i^N \sum_m P_i(m) \log \frac{P_i(m)}{Q_i(m)}$ and the Mean Chebyshev distance, $\overline{\mathrm{CD}} = \frac{1}{N}\sum_i^N \max_m |P_i(m) - Q_i(m)|$. Here, $P_i(m)$ refers to the occupancy of the $m^{th}$ state of the $i^{th}$ dihedral under distribution $P$. Table 1 provides an overview. For the Context-conditioned KMC, the mean KL was 0.02 (i.e., 2% information divergence) and the mean CD was 0.053 (showing a maximum mismatch of 5.3%). The corresponding values for the independent KMC were $\overline{KL} = 0.014$ and $\overline{CD} = 0.035$ . In case of the CC-Synced KMC, the mean KL divergence was 0.009 while the mean CD was 0.036.

Table 1. Quantitative comparison between KMC-derived populations and MD simulations, measured by mean divergence and total variation distance. The last row reports the baseline internal divergence of the MD data, computed across two equal halves of the trajectories.

| **Systems compared** | $\overline{\mathbf{KL}}$ | $\overline{\boldsymbol{CD}}$ |
|---|---|---|
| Independent KMC (Model A) vs. MD | 0.014 | 0.035 |
| Context-conditioned KMC (Model B) vs.MD | 0.020 | 0.053 |
| CC-Synced KMC (Model C) vs.MD | 0.009 | 0.036 |
| MD vs. MD | 0.043 | 0.067 |

To place these metrics in context, an internal sampling error analysis was conducted by splitting the MD trajectory into two equal independent blocks. The internal baseline divergence between the MD halves yielded a mean KL divergence of $0.043$ and a mean CD distance of $0.067$. Because the error between the CC-Synced KMC (Model C) and MD (mean

KL=$0.009$, mean $CD = 0.036$) is substantially smaller than the intrinsic statistical variation between independent MD subsets, we conclude that the CC-Synced KMC model (as well as the other KMC models) reproduces the MD equilibrium conformational ensemble well within statistical noise.

### 3.7 Comparing joint probability distributions

We investigate the backbone correlations captured in the KMC models. Figure 7 illustrates the joint probability distributions $P(\psi_{11}, \phi_{11})$ and conditional probabilities $P(\psi_{11}|\phi_{11})$ for the neighboring dihedrals of residue GLY11, comparing MD simulations against the three models. Here, $P(\psi_{11} \mid \phi_{11})$represents the probability of dihedral $\psi_{11}$ occupying a given state conditioned on the state of its neighbor $\phi_{11}$.

The MD-derived data reveals clear inter-dihedral correlations characteristic of Ramachandran-like preferences in discrete states shown in Figure 7 (a, b). For instance, when $\phi_{11}$ is in state 0, the conditional probabilities for states 0, 1, and 2 of $\psi_{11}$ are [$0.711$,0,$0.289$]. However, when $\phi_{11}$ shifts to state 1, these probabilities change to [0,0.15 ,$0.984$], indicating that the dominant state of $\psi_{11}$ switches from state 0 to state 2 as its neighbor flips. In contrast, the independent KMC model, shown in panels (c, d), fails to capture this dependency, displaying essentially uniform conditional distributions across all $\phi_{11}$ states.

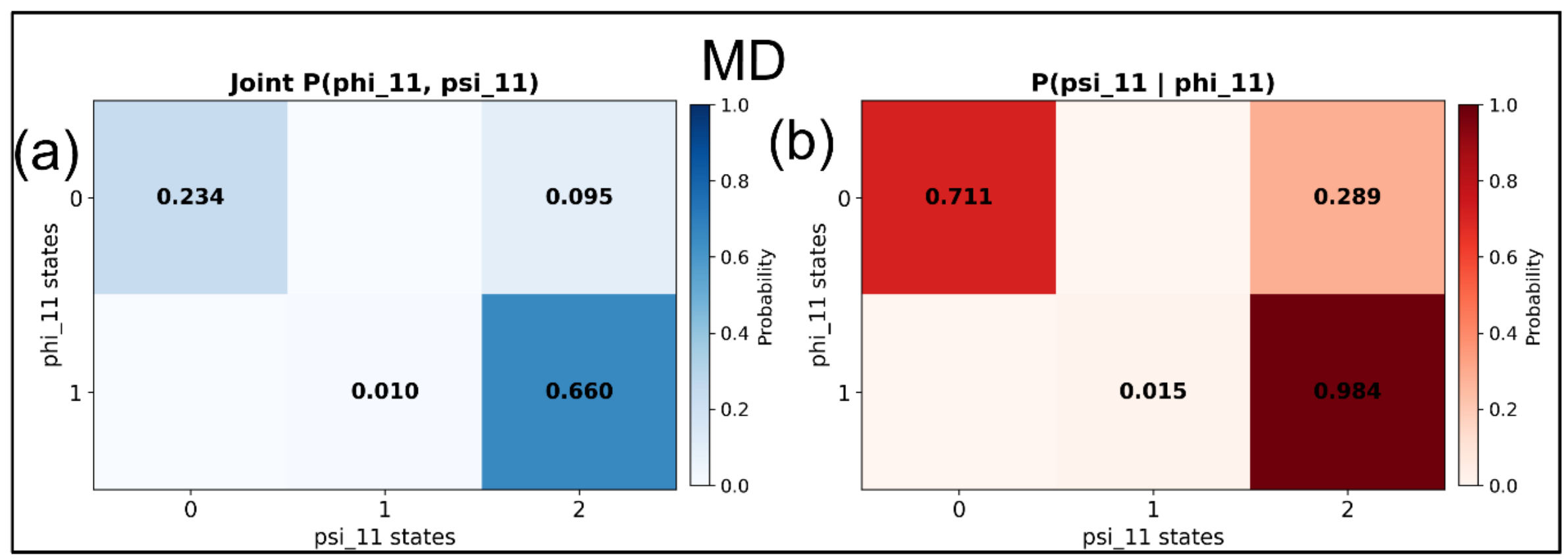
MD
(a)
Joint P(phi_11, psi_11)
0.234
0.095
0.010
0.660
phi_11 states
psi_11 states
Probability
(b)
P(psi_11 | phi_11)
0.711
0.289
0.015
0.984

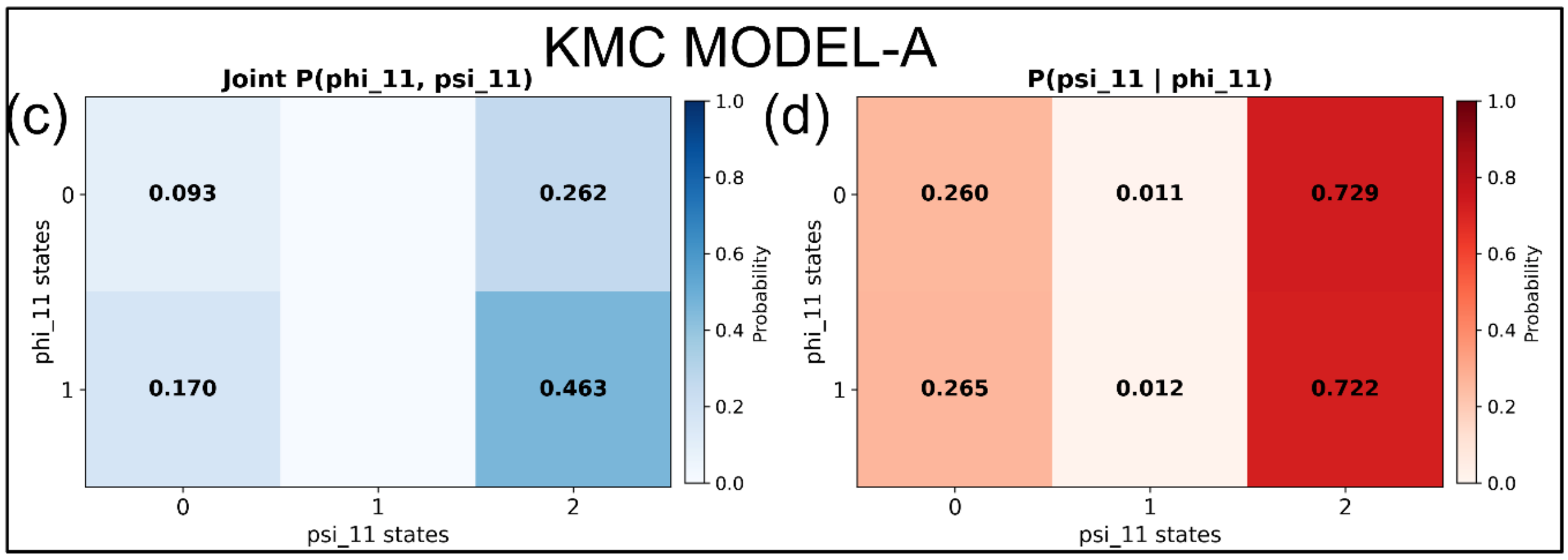
KMC MODEL-A
(c)
Joint P(phi_11, psi_11)
0.093
0.262
0.170
0.463
(d)
P(psi_11 | phi_11)
0.260
0.011
0.729
0.265
0.012
0.722

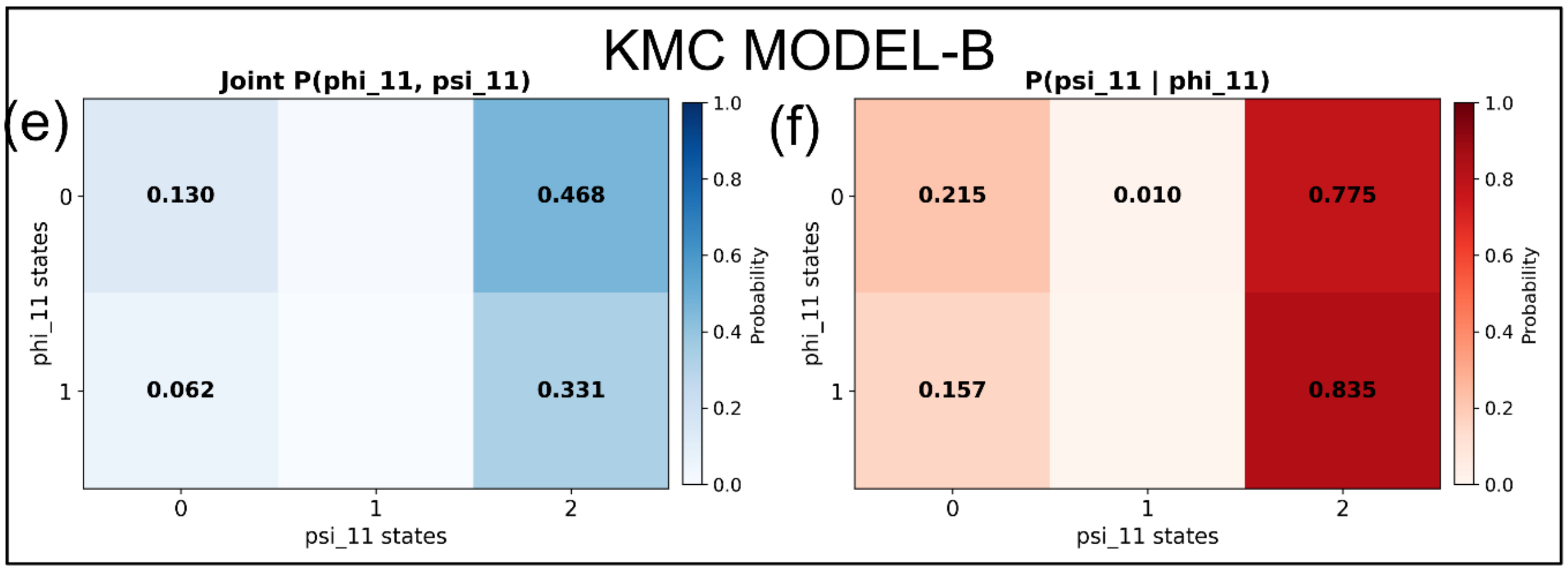
KMC MODEL-B
(e)
Joint P(phi_11, psi_11)
0.130
0.468
0.062
0.331
(f)
P(psi_11 | phi_11)
0.215
0.010
0.775
0.157
0.835

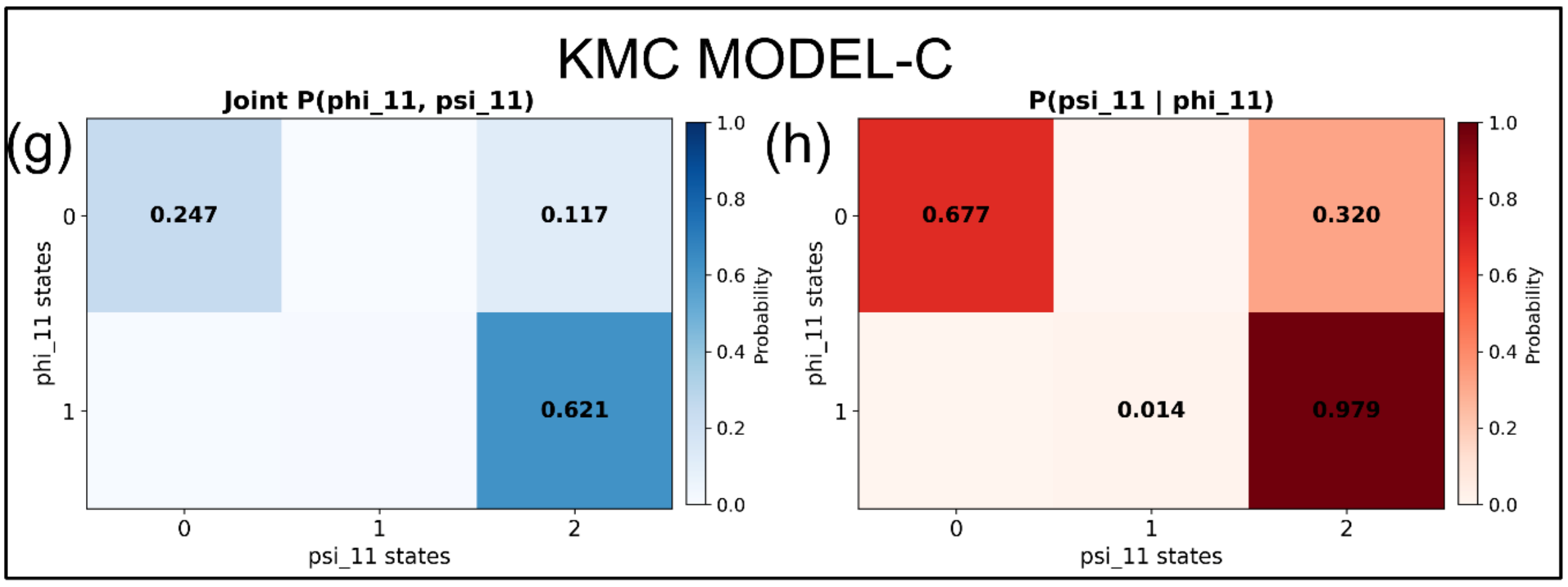
KMC MODEL-C
(g)
Joint P(phi_11, psi_11)
0.247
0.117
0.621
(h)
P(psi_11 | phi_11)
0.677
0.320
0.014
0.979

Figure 7. Probability distributions of the two dihedrals across different simulation models. (a),(c),(e) and (g) Joint Probability distributions $P(\phi_{11}, \psi_{11})$. (b),(d),(f) and (h) Conditional Probability distributions $P(\psi_{11}|\phi_{11})$. Probabilities less than $0.01$ are not mentioned.

While the Model B (e, f) captures the qualitative coupling trend observed in MD, its numerical agreement remains poor. Specifically, the conditional probabilities for $\psi_{11}$ shift from $[0.215,0,0.775]$ when $\phi_{11}$ is in state 0 to $[0.159,0,0.835]$ when $\phi_{11}$ is in state 1, showing significant deviations from MD when $\phi_{11}$ occupies state 0. Conversely, Model C, shown in panels (g, h**),** reproduces the MD conditional probabilities with remarkable accuracy ($[0.677,0,0.320]$ for $\phi_{11}$ state 0 and $[0,0.014,0.979]$ for state 1). Model C successfully propagates dihedral correlations as designed—a critical requirement for accurately modelling protein backbone dynamics and flexibility.

## 3.8 Correlation Map

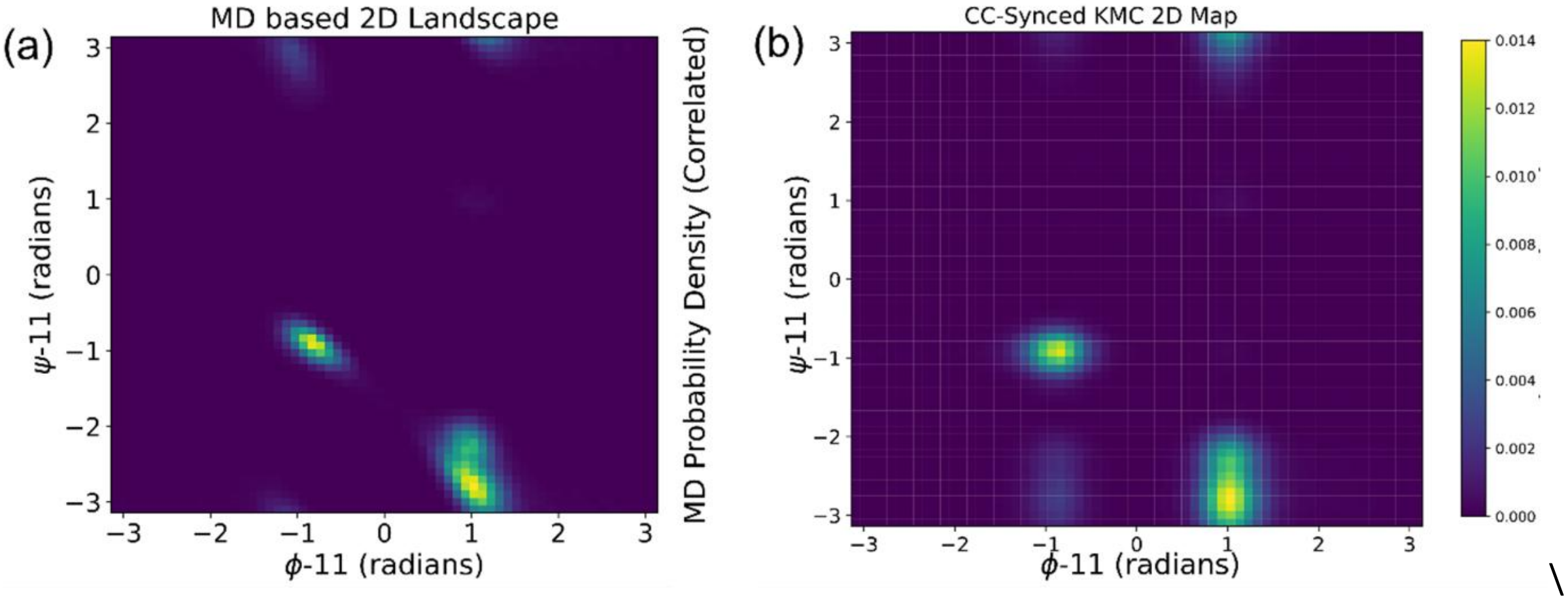

Figure 8. $\phi - \psi$ distribution maps predicted by MD (a) and the CC-Synced-KMC (b). The KMC based predicted map is obtained by combining the discrete state joint probability of $\phi_{11}$ and $\psi_{11}$ obtained from the KMC model with the marginal distributions of each dihedral for each state.

Finally, we calculate the continuous joint probability map by combining the macro-level behaviour of the system with the micro-level structural data. We use the discrete-state joint probability matrix to determine the global likelihood of each cross-state combination (eg. $\phi_{11}$ being in state $0$ while $\psi_{11}$ is in state 2). Next, we use the state-wise 1D histograms previously obtained from the training MD data and blend the discrete states with the continuous angular distributions to get the KMC predicted $\phi - \psi$ map for GLY11 shown in Figure 8(b) which exhibits strong agreement with the reference MD-derived map. Figure 8 is one of the key results from this work. This strong correspondence highlights the success of the Context-conditioned Synced KMC model in capturing complex conformational correlations without losing single-dihedral fidelity.

Table2. The total unique conformational state space and the total number of dihedral moves obtained in the three methods.

| | Cumulative trajectory time ($\mu s$) | Number of Global Conformations | Total Dihedral Flips |
|---|---|---|---|
| MD | 7.3 | 1655 | 17,338 |
| KMC Model A | 100 | 17667 | 235,744 |
| Model B KMC Model B | 100 | 17108 | 243,378 |
| KMC Model C | 100 | 13845 | 242,536 |
| KMC Model C | $10^6$ | $> 10^8$ | $> 10^9$ |

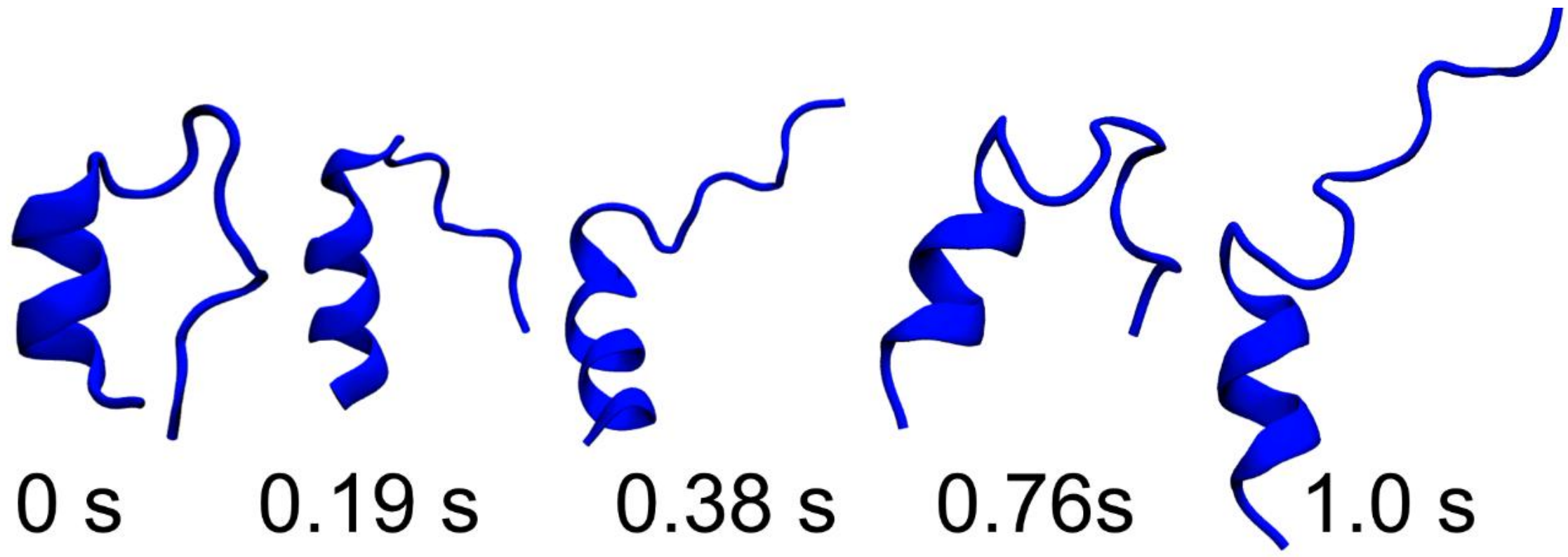

Figure 9. Snapshots from a single $1\ second$ trajectory using KMC Model C.

### 3.9 Performance metrics

A quantitative comparison of the three simulation protocols is presented in Table 2. The combined MD trajectory, totaling 7.3 µs, sampled 1,655 unique conformational states. In contrast, the KMC implementations achieved a significantly more extensive mapping of the conformational landscape. Models A and B report more than 17,000 global states while Model C shows a slight shrinkage to $\sim$13,000 from $100\mu s$ trajectory time. From a computational efficiency standpoint, KMC outperforms standard MD by several orders of magnitude, executing the 100 µs simulations in less than two minutes on a desktop computer.

In addition, a single trajectory of length $1\,second$ was generated using CC-Synced KMC (Model C) in 14 hours on a single core processor. More than 2.4 billion moves were executed in the trajectory. Snapshots from the $1\,s$ trajectory are presented in Figure 9. These results underscore the utility of KMC as an agile framework for rapid ensemble generation, leveraging MD for parameterization rather than brute-force sampling.

## 4. Conclusions

In this work, we introduced a dihedral-based kinetic Monte Carlo (KMC) framework that accelerates protein dynamics over extended timescales while retaining microscopic resolution. By parameterizing dihedral flip rates from short molecular dynamics (MD) trajectories, we evaluated three KMC models of increasing complexity. While all models accurately reproduced the MD equilibrium ensemble, only the Context-Conditioned Synced KMC (CC-Synced KMC) model successfully captured local dynamic correlations and concerted transitions between neighboring dihedrals. By bridging short-time MD parameterization with rapid KMC state-space exploration, this method achieves fine torsional resolution at a fraction of the computational cost of conventional MD.

The fundamental strength of this approach lies in defining state space through local dihedral coordinates rather than global collective variables or reaction coordinates. This modular design circumvents the combinatorial explosion of states and memory bottlenecks that typically plague kinetic network construction for larger molecules. Consequently, the framework eliminates the need for brute-force MD sampling across the entire kinetic network. We believe that by constructing global kinetics from localized torsional transitions, the method can predict global conformational states never explicitly observed in the underlying MD trajectories.

Finally, this framework addresses a long-standing challenge in computational structural biology: the trade-off between spatial resolution and temporal reach. By shifting MD from a brute-force sampling tool into a high-throughput parameterization engine for CC-Synced KMC, the method retains critical physical couplings without the prohibitive cost of explicit force calculations. While demonstrated here on backbone $\phi, \psi$ dihedrals, the protocol naturally extends to side-chain rotamers, offering a scalable blueprint to map multi-state energy landscapes, identify rare transient intermediates, and characterize long-timescale dynamics in complex protein systems. As demonstrated in the example system, the seconds timescale is within reach without sacrificing microscopic resolution.

## Author Contributions

SB and AC contributed to conceptualization and writing the manuscript. The coding and analysis were performed by SB. The simulations were performed by RD and GT.

## Supporting Information

Additional methodological details, parameterization protocols for KMC Models A–C, supplementary figures S1–S6, and summary tables are provided in the Supporting Information.

## Funding

SB acknowledges the SERB (ANRF) grant CRG/2022/004108.

## Disclosure Statement

The authors confirm that there are no known conflicts of interest associated with this publication.